\documentclass[aps,nofootinbib,nobibnotes,noeprint,preprintnumbers,floatfix,twocolumn]{revtex4-2}
\usepackage{graphicx} 
\usepackage{hyperref}
\usepackage[utf8]{inputenc}
\usepackage{graphicx}
\usepackage{amsmath}
\usepackage{xcolor}
\usepackage[caption = false]{subfig}
\usepackage{graphicx}
\usepackage{dcolumn}
\usepackage{bm}
\usepackage{hyperref}
\usepackage{xcolor}
\usepackage{amsmath,amsfonts,amssymb}
\usepackage{commath}
\usepackage{mathtools}
\usepackage{tikz}
\usepackage[utf8]{inputenc} 
\usepackage[T1]{fontenc}

\def\aap{{A\&A}}

\newcommand{\bea}{\begin{eqnarray}\begin{aligned}}
\newcommand{\eea}{\end{aligned}\end{eqnarray}}

\newcommand{\alphaem}{\alpha_\text{em}}

\newcommand{\cm}{\text{cm}}

\newcommand{\eV}{\text{eV}}
\newcommand{\kev}{\text{keV}}
\newcommand{\gev}{\text{GeV}}

\newcommand{\Eq}[1]{Eq.~(\ref{eq:#1})}

\newcommand{\CMB}{\text{CMB}}

\newcommand{\BH}{{\text{BH}}}

\newcommand{\halo}{\text{halo}}

\newcommand{\fa}{f_a}

\newcommand{\polar}{\text{polar}}
\newcommand{\ga}{g_{a\gamma\gamma}}

\newcommand{\grad}{\text{grad}}

\newcommand{\tot}{\text{tot}}

\newcommand{\NFW}{\text{NFW}}

\newcommand{\Sec}[1]{Sec.~\ref{sec:#1}}

\newcommand{\Appx}[1]{Appendix~\ref{appx:#1}}

\newcommand{\Fig}[1]{Fig.~\ref{fig:#1}}

\newcommand{\decay}{\text{decay}}

\newcommand{\Thetazero}{\Theta_0}

\newcommand{\osc}{\text{osc}}

\newcommand{\mass}{\text{mass}}

\newcommand{\grav}{\text{grav}}

\newcommand{\gr}{\text{gr}}

\begin{document}
\preprint{FERMILAB-PUB-23-635-T}

\title{Detecting Axion Dark Matter with Black Hole Polarimetry}
\author{Xucheng Gan$^1$}
\email{xg767@nyu.edu}
\author{Lian-Tao Wang$^{2,3,4}$}
\email{liantaow@uchicago.edu}
\author{Huangyu Xiao$^{4,5}$}
\email{huangyu@fnal.gov}

\affiliation{$^1$Center for Cosmology and Particle Physics, Department of Physics, New York University, New York, NY 10003, USA}

\affiliation{$^2$ Department of Physics, The University of Chicago, Chicago, IL 60637, USA}
\affiliation{$^3$ Enrico Fermi Institute, The University of Chicago, Chicago, IL 60637, USA}
\affiliation{$^4$Kavli Institute for Cosmological Physics, University of Chicago, Chicago, IL 60637, USA}
\affiliation{$^5$ Astrophysics Theory Department, Theory Division, Fermilab, Batavia, IL 60510, USA}

\begin{abstract}
The axion, as a leading dark matter candidate, is the target of many ongoing and proposed experimental searches based on its coupling to photons. Ultralight axions that couple to photons can also cause polarization rotation of light, which the cosmic microwave background can probe. In this work, we show that a large axion field is inevitably developed around supermassive black holes due to the Bose-Einstein condensation of axions, enhancing the induced birefringence effects. Therefore, measuring the modulations of supermassive black hole imaging polarization angles is a strong probe of the axion-photon coupling due to the formation of the axion condensation~(axion star) which enhances the axion field. The oscillating axion field around black holes induces polarization rotation on the black hole image, which is detectable and distinguishable from astrophysical effects on the polarization angle, as it exhibits distinctive temporal variability and frequency invariability. In this work, we perform theoretical calculations of the axion star formation rate and the corresponding enhanced axion field value near supermassive black holes. Then, we present the range of axion-photon couplings within the axion mass range $10^{-21}\text{--}10^{-16}\,\eV$ that can be probed by the Event Horizon Telescope. The axion parameter space probed by black hole polarimetry will expand with improvements in sensitivity of polarization measurements and more black hole polarimetry targets with determined black hole masses.
\end{abstract}
\maketitle
\section{Introduction}

\begin{figure*}[t]
\centering
\includegraphics[width=1.1\columnwidth]{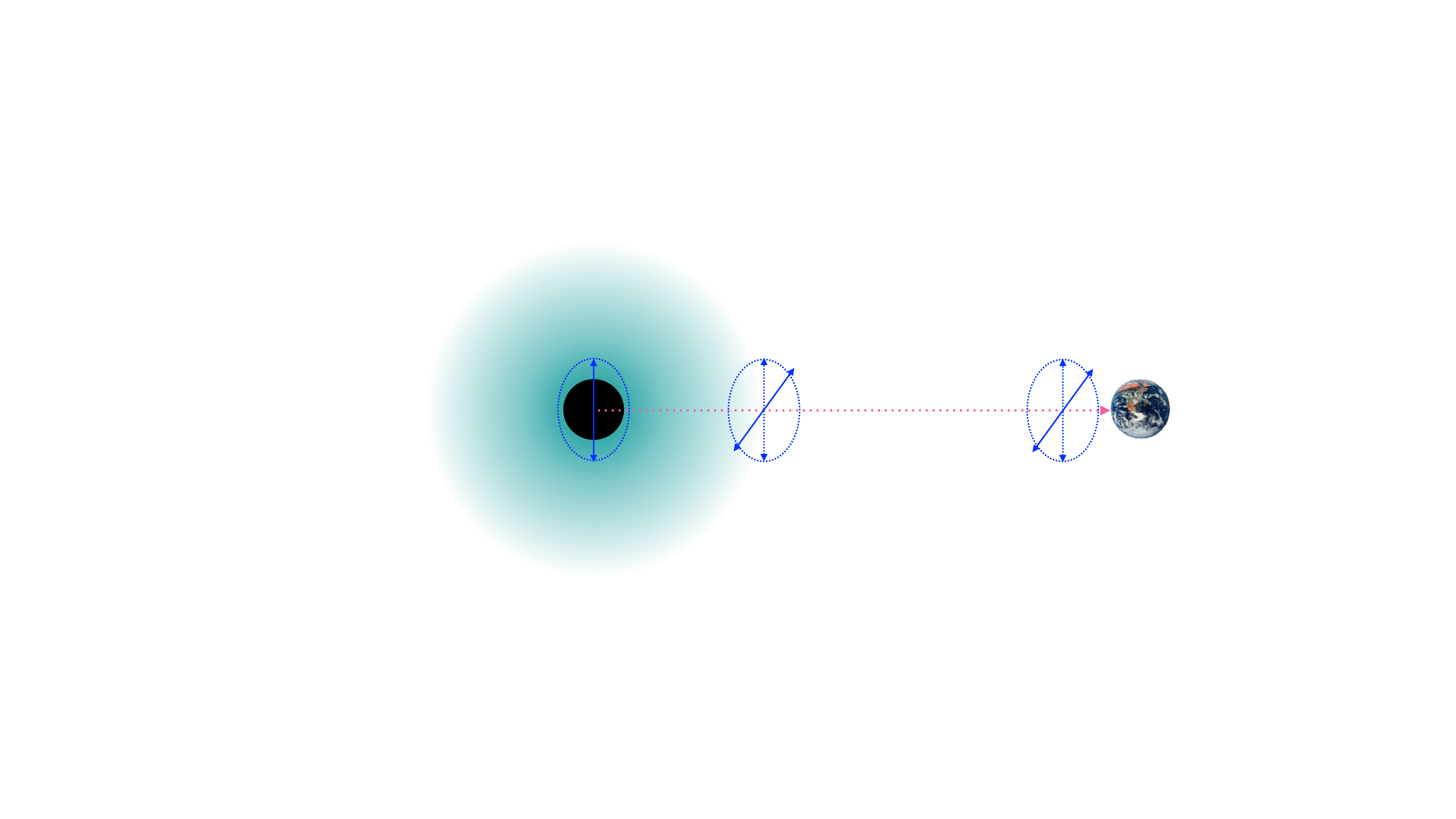}
\caption{The schematic diagram of the polarimetry of supermassive black hole~(SMBH) with the axion star surrounding it. In this diagram, the small black circle denotes the SMBH, the large cyan circular cloud denotes the axion star in SMBH's gravitational background, and $R_*$ denotes the radius of the axion star. When the light produced near the vicinity of the SMBH~(orange dotted line) propagates through the axion star to Earth, the light's polarization angle~(blue solid arrowed lines) is rotated by the oscillating axion field with a rotation angle $\Delta\phi_\polar$. Such variation on the polarization also has the time modulation with an angular frequency $\omega_a \sim m_a$, and the modulation is coherent over the size of the axion star.}
\label{fig:axion_polar}
\end{figure*}
The axion, proposed to solve a fundamental puzzle in strong interactions, is also a viable dark matter candidate \cite{Peccei:1977hh,Weinberg:1977ma,Wilczek:1977pj,Kim:1979if,Abbott:1982af,Dine:1982ah,Preskill:1982cy, Peccei:2006as}. The current spectrum of axions expands to models that do not solve the puzzle, which motivates us to consider a broader range of axion parameters~\cite{Arvanitaki:2009fg}. 
Laboratory searches for axions in the ultralight mass ranges ($m_a\lesssim 10^{-11}\,\eV$) have also been proposed and performed.  These experimental methods include magnetometers~\cite{Sikivie:2013laa,Kahn:2016aff,Ouellet:2018beu,Gramolin:2020ict,Salemi:2021gck,Zhang:2021bpa,DMRadio:2022jfv}, cavities~\cite{Jaeckel:2007ch,ALPS:2009des,Caspers:2009cj,Ehret:2010mh,Redondo:2010dp,Bahre:2013ywa,Betz:2013dza,OSQAR:2015qdv,Janish:2019dpr,Berlin:2019ahk,Berlin:2020vrk,Gao:2020anb,Berlin:2022hfx}, and optical interferometers~\cite{DeRocco:2018jwe,Obata:2018vvr,Liu:2018icu,Fedderke:2023dwj}. On the other hand, astrophysical and cosmological observations placed the strongest bound on axion parameters in this mass range, such as the spinning down of black holes from superradiance~\cite{Baryakhtar:2020gao,Mehta:2020kwu,Unal:2020jiy}, recurrent axinovae~\cite{Fox:2023aat}, spectral distortion~\cite{Mirizzi:2005ng, Mirizzi:2009nq,Tashiro:2013yea,Mukherjee:2018oeb,Chang:2023quo}, X-ray observations~\cite{Wouters:2013hua,Marsh:2017yvc,Reynolds:2019uqt,Dessert:2019sgw,Buschmann:2019pfp,Dessert:2020lil,Dessert:2021bkv,Reynes:2021bpe,Nguyen:2023czp}, gamma-ray observations~\cite{Mirizzi:2009aj,Horns:2012pp,Meyer:2013pny,Fermi-LAT:2016nkz,Meyer:2016wrm,Meyer:2020vzy,Mastrototaro:2022kpt}, solar basin~\cite{VanTilburg:2020jvl,DeRocco:2022jyq}, heating of dwarf galaxies~\cite{Dalal:2022rmp,Wadekar:2021qae, Wadekar:2022ymq}, cosmic axion background~\cite{Dror:2021nyr,Langhoff:2022bij,ADMX:2023rsk}, finite density effects inside stellars~\cite{Hook:2017psm,Balkin:2022qer}, radio telescope~\cite{Huang:2018lxq,Caputo:2018ljp,Safdi:2018oeu,Caputo:2018vmy,Sun:2021oqp,Buen-Abad:2021qvj,Sun:2023gic}, and solar telescope~\cite{Raffelt:1985nk,CAST:2017uph,OHare:2020wum}.
In this work, we propose a complementary search on axions using measurements of the modulation of the electric vector position angle~(EVPA) of black hole images. The polarization angle in the vicinity of black holes will experience birefringence effects from the dense axion field, which is developed through axion dark matter accretion onto black holes. We illustrate this concept schematically in \Fig{axion_polar}.

The axion field with masses $\lesssim 10^{-18} \,\eV$, if accounting for the dark matter density, can rotate the polarization of light passing through it, which is known as the cosmic birefringence effect. It can be probed in cosmic microwave background (CMB) polarization measurements~\cite{Harari:1992ea,Fedderke:2019ajk,Diego-Palazuelos:2022dsq,Luo:2023cxo}. Extensive searches for axionlike polarizations in CMB have recently been performed by BICEP/Keck~\cite{BICEPKeck:2021sbt}, POLARBEAR~\cite{POLARBEAR:2023ric}, and SPT-3G~\cite{SPT-3G:2022ods}. Other polarimetric observations that can detect axion strings or axion dark matter are studied in \cite{Agrawal:2019lkr,Liu:2019brz,Yuan:2020xui,Jain:2021shf,Yin:2021kmx,Liu:2021zlt,Castillo:2022zfl,Yao:2022col,Hagimoto:2023tqm,Fortin:2023jlg}.
In our proposal, the axion field value is greatly enhanced compared to its expectation value predicted by the cosmic dark matter density due to the formation of axion stars in the gravitational potential well of black holes, increasing our sensitivity to axion-photon couplings. 
The first image of supermassive black holes~(SMBHs) M87$^*$ by the Event Horizon Telescope~(EHT)~\cite{EventHorizonTelescope:2019pgp} opens the window into the new physics beyond the standard model that is only accessible around the Schwarzschild radius of black holes. Our axion stars near the supermassive black holes mostly have a size much larger than the Schwarzschild radius; therefore, it does not necessarily require such high spatial resolution.
The polarimetric studies of Event Horizon Telescope targets have been performed actively with interferometric observations \cite{ALMA:2021axn, EventHorizonTelescope:2021srq, EHT:2024nwx}, further enabling us to use black hole polarimetry to study the birefringence effects induced by axion stars with larger radii.

Therefore, the crucial question becomes whether the axion star formation rate is large enough to build up a large axion field value around the SMBH. Assuming the axion is the dark matter, the self-scattering of ultralight axions is greatly enhanced due to the large phase density. Both numerical and analytical studies confirmed the formation of axion stars (sometimes called solitons) in dense dark matter halos from the gravitational scattering of axion waves or from the quartic self-couplings \cite{Kolb:1993zz,Schive:2014hza,Levkov:2018kau,Eggemeier:2019jsu,Chen:2021oot,Chan:2022bkz,Glennon:2023jsp,Jain:2023tsr}, which provide us with the tools to study the axion field value after the axion star formed. In current studies of axion stars, stable configurations of axion fields are numerically solved, with the dilute branch of axion stars being balanced by their gravity and kinetic pressure, while the dense branch is unstable due to the emission of relativistic axions \cite{Eby:2015hyx,Visinelli:2017ooc,Eby:2019ntd}. We obtain a solution of stable axion field configurations in the black hole background, which allows a much larger stabilized axion field than the usual dilute branch. 
Stable axion stars around black holes will provide a significant and unique photon signal with a modulating polarization angle. 
In this work, we focus on the theoretical calculation of the axion star formation rates around supermassive black holes, which enhance axion field values. Similar ideas were proposed in the context of the axion cloud generated by black hole superradiance \cite{Chen:2019fsq,Chen:2021lvo,Chen:2022oad,Shakeri:2022usk}. Our calculation suggests that a broader axion mass range can be probed with individual black holes since we do not rely on the axion cloud generated by BH superradiance but the axion cloud generated by accretion around the SMBH. Our work demonstrates the great potential of using black hole polarimetry as detectors of axion dark matter, which will motivate both observational efforts on supermassive black holes and theoretical progress on the exact axion star formation rate in the black hole background.

\section{Axion Stars around black holes}\label{sec:AS_BH_solution}

In the following, we will study the solution of axion stars around black holes and the formation rate of such objects in the limit that the black hole is dominating the gravitational interaction, which will enable us to calculate the axion field value.
Solving axion electrodynamics in the presence of axion fields, one would obtain modified Maxwell's equations under which photons with different polarizations will propagate differently. The polarization angle shift induced by the axion field is \cite{Harari:1992ea,Fedderke:2019ajk}
\begin{equation}
\label{eq:phi_polar_calc}
    \Delta\phi_{\rm polar} = \frac{\ga}{2}[a_{\rm source}-a_{\rm earth}],
\end{equation}
where $a_{\rm source}$ and $a_{\rm earth}$ are the axion field values at source and earth respectively. They are related to the energy density of axions $\rho_a$ and the axion mass $m_a$ and are written as
\begin{equation}
a = \sqrt{2\rho_a/m_a^2} \, \cos(m_a t),
\end{equation}
which reveals the characteristics of time modulations with the frequency $m_a$. Therefore, if there is a significant difference in the axion amplitude between the source and Earth, the polarization angle shift can be large. It is crucial to calculate the axion field value developed around black holes to determine the exact value of the polarization angle induced by axions. We will study a new solution of stable axion stars around black holes and the formation rate of those axion stars to determine the axion field value.

\subsection{Stable Axion Field Configuration Around Black Holes}
The axion Lagrangian we consider in this work is 
\bea
\mathcal{L} = - \frac{1}{4} F^2 + \frac{1}{4} \ga a F \widetilde{F} + \frac{1}{2}\partial_\mu a \partial^{\mu} a  - V(a),
\eea
where the axion potential can be expressed as the quartic self-coupling and the mass term 
\bea
V(a) \simeq \frac{1}{2} m_a^2 a^2 - \frac{\lambda}{4!} a^4.
\label{eq:VaGeneric}
\eea
The quartic coupling can be related to the axion decay constant as $\lambda=m_a^2/\fa^2$. Axion-photon couplings can be parametrized as $\ga = c_\gamma \alphaem/(2 \pi \fa)$, where $c_\gamma$ is a model-dependent constant. Although typically $c_\gamma \sim 1$, it could be easily enhanced in various models, including the clockwork axion model~\cite{Farina:2016tgd,Agrawal:2017cmd,Dror:2020zru}, the $\mathbb{Z}_N$ models~\cite{Hook:2018jle,Dror:2020zru,DiLuzio:2021pxd}, and the two-axion alignment models~\cite{Agrawal:2017cmd}. Therefore, we will treat $\ga$ as a free parameter and study the phenomenology of axion stars with axion-photon couplings. In the following discussion, we will see that the axion with $c_\gamma \sim 1$ can be detected by the next-generation Event Horizon Telescope~(ngEHT), and the axion with $c_\gamma \sim 10^2$ can be detected by the current EHT observation.

We will study the axion star solution under the assumption $M_\BH \gg M_*$, where $M_*$ is the axion star's mass. 
For the moment we neglect the axion self-interactions arising from the axion potential, which will appear later when we calculate the critical mass of axion stars above which axion fields cannot remain stable. To describe the axion's profile in the background of the black hole quantitatively, we solve the Klein-Gordon equation
\bea
\label{eq:a_KG_Eq}
\nabla_\mu \nabla^\mu a = - V'(a)
\eea
in the black hole background, where $\nabla_\mu$ is the covariant derivative in the curved background of the black hole. To solve the time-dependent profile of axion, we write the axion profile as
\bea
\label{eq:axion_profile}
a(r,t) = \fa \,\Theta(r)\cos(\omega_a t),
\eea
where $\Theta(r)$ is the spatial-dependent part of the oscillation angle. The Schr\"{o}dinger-Newton-type equation that governs the spatial part of the axion fields is written as
\begin{equation}
\label{eq:schrodinger}
\pmb{\nabla}^2 \Theta = 2 \left(m^2_a \Phi+\frac{m^2_a -\omega^2_a}{2} \right) \Theta,
\end{equation}
where 
\bea
\label{eq:Phi_BH_main}
\Phi(r) \simeq - \frac{G M_\BH}{r}
\eea
is the gravitational potential of the black hole. Here, we have made the assumption $R_* \gg G M_{\rm BH}$; therefore, the black hole can be approximated by a pointlike source. Because we focus on the region where $M_\BH \gg M_*$, we ignore the axion's back reaction to the BH background. The contribution of the gravitational field of the BH  has not been included in previous studies but is essential in this work to generate a significantly enhanced axion field in stable axion stars.

Noting that \Eq{schrodinger} has the same form as the Schr\"{o}dinger equation for the electron radial wave function of hydrogen, we obtain the eigenfunction
\bea
\label{eq:schrodinger_BH_sol}
\Theta(r) = \Thetazero \, e^{-G M_\BH m_a^2 r},
\eea
and the eigenfrequency
\bea
\label{eq:schrodinger_BH_omega}
\omega_a = m_a \sqrt{1 - \left( G M_\BH m_a \right)^2 }.
\eea
$\Theta(r)$ in \Eq{schrodinger_BH_sol} is the profile of the axion star in the ground state, which is spherically symmetric. $\omega_a$ in \Eq{schrodinger_BH_omega} is the axion's oscillation frequency given the ground state profile. The coefficient $\Thetazero$ represents $\Theta(r=0)$ and it needs to be determined by the amount of mass accreted onto the gravitational potential well of black holes. The above axion profile has interesting implications for the polarization measurements. It suggests that the stable configuration of the axion field will have a spherical profile that oscillates coherently over space with the same frequency that is determined by the axion mass. Therefore, axion-induced polarization angles exhibit both spatial correlations and oscillatory temporal variations, which can be distinguished from contributions from other astrophysical sources. On the other hand, astrophysical sources might also cause a periodic rotating polarization angle. The observed rotation of the electric vector position angle at a timescale of $\sim$ 70 min of the Galactic Center SMBH Sgr A$^*$ has been interpreted as the orbital motion of a hot spot embedded in a magnetic field \cite{Wielgus:2022heh}. Looking at the spatially correlated signals should suppress astrophysical systematics, which we leave for future explorations. A similar discussion on the ultralight particle in the gravitational background,  referred to as the ``gravitational atom'', can be found in \cite{Banerjee:2019epw,Chavanis:2019bnu,Banerjee:2019xuy,Budker:2023sex}. From \Eq{schrodinger_BH_sol}, we find that the axion star radius is
\bea
\label{eq:R_star}
R_* \sim \frac{1}{G M_\BH m_a^2},
\eea
which is independent of the mass of the axion star. This is distinctly different from the mass-radius relation of self-gravitating axion stars which obeys $R_*^\text{self} \sim 1/(G M_* m_a^2)$. Such difference can be explained by the difference in the gravitational energy of these two kinds of axion stars. For stable configurations, the gradient energy balances the gravitational energy, and therefore, the radii of these two axion stars have different $M_*$-dependence. 

To estimate the axion star's mass, we use the formula
\bea
\label{eq:M_star_eq_approx}
M_* = \int d^3 \pmb{r} \, \rho_*  \sim \rho_* R_*^3,
\eea
where $\rho_*$ is the energy density of the axion star. We have
\bea
\label{eq:rho_star}
\rho_* \simeq \Big\langle \frac{1}{2} \dot{a}^2 + \frac{1}{2} \nabla a^2 + \frac{1}{2} m_a^2 a^2 - \frac{\lambda}{4!} a^4 \Big\rangle,
\eea
where $\langle \cdots \rangle$ represents the time-averaged values, the first term is the oscillation energy, the second term is the gradient energy, the third term is the potential energy from the axion's mass term, and the fourth term is the potential energy from the axion's self-interaction. From \Eq{schrodinger_BH_sol}, we find that the gradient energy is suppressed by an extra $(G M_\BH m_a)^2$ factor, and therefore it is negligible when $G M_\BH m_a \ll 1$. Before the axion star's mass reaches the critical value, the energy contribution from the self-interaction can also be neglected. Therefore, the axion star's energy density can be approximately written as 
\bea
\label{eq:rho_star_approx}
\rho_* \sim m_a^2 f_a^2 \Thetazero^2. 
\eea
Substituting \Eq{R_star} and \Eq{rho_star_approx} into \Eq{M_star_eq_approx}, we have
\bea
\label{eq:M_star_result_approx}
M_* \sim \frac{f_a^2 \Thetazero^2}{G^3 M_\BH^3 m_a^4 },
\eea
which reveals the relation between $M_*$ and $\Thetazero$. From \Eq{phi_polar_calc}, we have
\bea
\label{eq:theta_polar_approx_0}
\abs{\Delta\phi_\polar} \simeq \frac{1}{2}\ga \Thetazero \fa, 
\eea
where $\abs{\cdots}$ denotes the amplitude of oscillation. Combining \Eq{M_star_result_approx} and \Eq{theta_polar_approx_0}, we can write the amplitude of  polarization angle as
\bea
\label{eq:theta_polar_approx}
\abs{\Delta \phi_\polar} \sim \ga \left(G^3 M_\BH^3 m_a^4 M_*\right)^{1/2},
\eea
which depends on the mass of the axion star. 

With self-interaction, only axion stars with mass smaller than some critical value $M_* < M_*^{\rm max}$ are stable against a bosenova \cite{Eby:2016cnq,Levkov:2016rkk}. Otherwise, the axion star will radiate the relativistic axions until its mass falls below the critical mass. In what follows, we will calculate the critical mass $M_*^{\max}$ of the axion star in the BH background. In the total energy of the axion star, the relevant components are the gradient energy, $E_*^\grad$, responsible for the kinetic pressure that stabilizes axion stars, and the attractive self-energy, $E_*^\lambda$, that causes a bosenova. These energy components of axion stars can be written as
\bea
\label{eq:E_star_lambda}
E_*^\lambda \simeq -\frac{\lambda G^3 m_a^2 M_\BH^3 M_*^2}{128 \pi}
\eea
and
\bea
\label{eq:M_star_grad}
E_*^\grad  \simeq \frac{1}{2}G^2 m_a^2 M_\BH^2 M_*.
\eea
The exact coefficients of these energy components are calculated in \Appx{axion_star_energy}.
The critical mass is reached when$
\abs{\frac{E_*^\lambda}{E_*^\grad}}  = \frac{1}{6}.$
This ratio is obtained by minimizing the total energy of axion stars and looking for viable solutions of mass-radius relation. See more discussions in \Appx{critical mass}. Hence, we obtain
\bea
\label{eq:M_star_max_2}
M_*^{\max} =\frac{32 \pi}{3 \lambda G M_\BH }.
\eea
The critical mass can increase if the self-coupling of axions is smaller. Since a larger axion star mass corresponds to a larger axion field value, the maximum $f_a$ with a critical star formation will achieve the largest field value induced by axion stars. The maximum polarization angle one can achieve with the maximum allowed axion star is thus given by
\bea
\label{eq:phi_polar_max_approx}
\abs{\Delta\phi_{\polar}^{\max}} \sim 2 \ga f_a  G M_\BH m_a.
\eea
Note that a large $f_a$ will correspond to a better sensitivity to the probe of axion-photon coupling $\ga$ since a larger critical star mass can be achieved. We will discuss what is the largest $f_a$ which still allows the formation of axion stars at the critical mass later. Substituting $\ga = c_\gamma \alphaem/(2\pi \fa)$ into \Eq{phi_polar_max_approx}, the maximum axion-induced rotation angle is $|\Delta \phi^{\max}_\polar| \sim c_\gamma \alphaem G M_\BH m_a$. From \Sec{sensitivity}, we know that the sensitivities of EHT and ngEHT are $3^\circ$ and $0.05^\circ$, respectively. Therefore, we know that the current EHT observation can test the axion with $c_\gamma \sim 10^2$ and the ngEHT can test the axion with $c_\gamma \sim 1$.

\begin{figure*}[t]
\centering
\includegraphics[width=1.29\columnwidth]{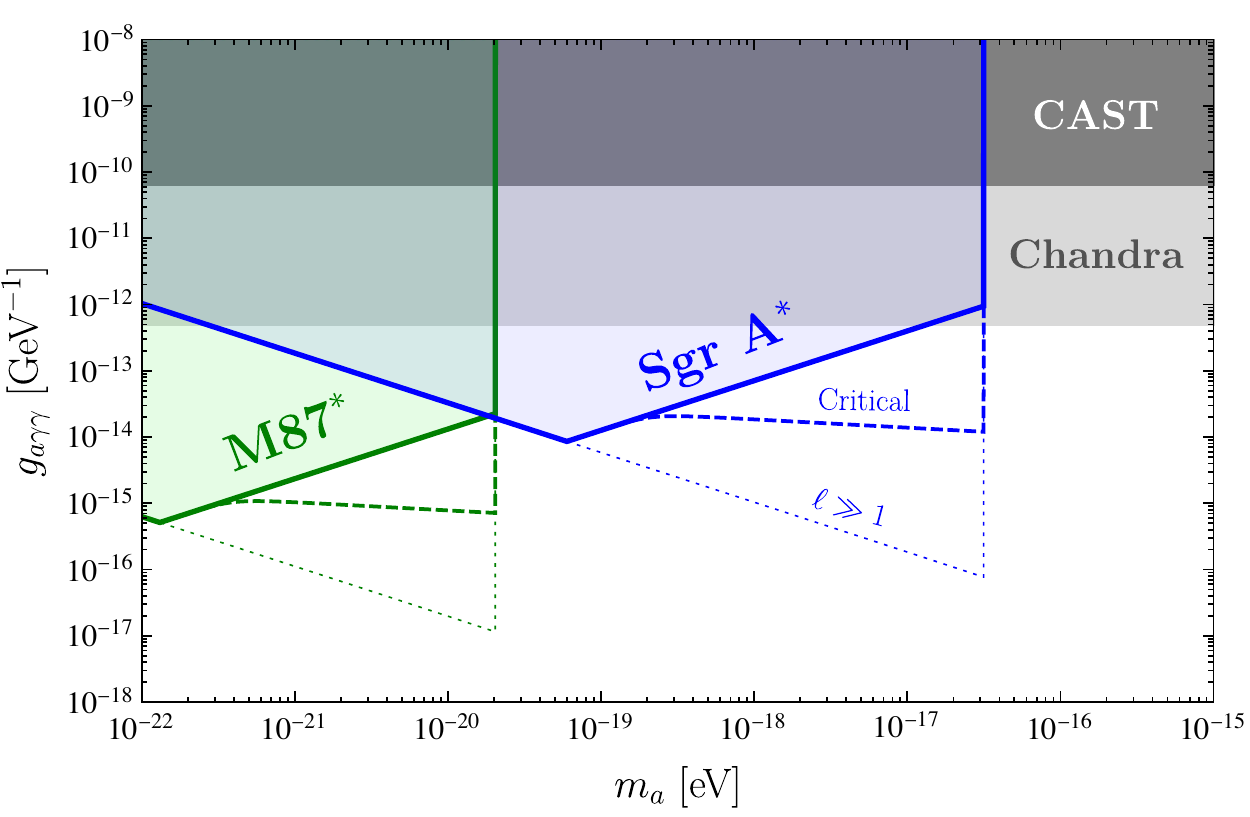}
\caption{The range of axion-photon couplings that can be probed by the birefringence effect of axion stars in black hole images. Here we use M87$^*$ and Sgr A$^*$ to discuss the detection capabilities, taking the precision of polarimetric measurement as $3^\circ$ for both of them. The masses of M87$^*$ and Sgr A$^*$ are taken to be $6.5\times10^9M_{\odot}$ \cite{EventHorizonTelescope:2019ggy} and $4.2\times 10^6M_{\odot}$ \cite{2019A&A...625L..10G}, respectively. The shaded parameter space shows the sensitivity by considering axion star formation from gravity only. In the region where $m_a$ is relatively small, $t_H<t_\decay$ and the solid lines obey $\ga \propto m_a^{-3/4}$. In the region where $m_a$ is large, $t_H>t_\decay$, from which we have $\ga\propto m_a^{3/4}$. The dashed ``critical'' curve encloses the parameter space expanded by including the formation rate from self-interactions where $f_a$ is chosen to form critical stars that maximize $\Delta \phi_\text{polar}$. The parameter space above the dotted curve shows the potential maximal sensitivity when the decay rate of axion stars $\Gamma_\decay \propto (G M_\BH m_a)^{5+4\ell}$ is suppressed for excited states with large $\ell$. }
\label{fig:ma_ga_plt}.
\end{figure*}

\subsection{Axion Star Formation at Halo Center}\label{sec:axion_form_halo_center}
Axions, originally unbound from each other, can form axion stars near black holes in the presence of self-scattering in dense environments. The consequence of axion star formation will populate dense axion clouds around supermassive black holes as gravitational atoms since the SMBH will accrete and absorb the axion star formed in the Galactic Center \cite{Davies:2019wgi}. Such an alignment between the axion star and the SMBH can be easily achieved because the SMBH has a much smaller potential, which maximizes the probability for the axion star to form around it. In our work, the enhancement in the axion star formation rate from BHs is not considered.
In the following, we will study the formation rate of axion stars in Navarro–Frenk–White~(NFW) halos and determine the axion field value. The relevant timescale in this problem is the scattering timescale between axions as wavelike particles, which approximately gives the condensation timescale of axion stars in simulations of axion fields. If we consider both gravity and self-interaction, the condensation timescale is \cite{Chen:2021oot}
\begin{equation}
\label{eq:tau}
\tau=\frac{\tau_{\rm self}\tau_{\gr}}{\tau_{\gr}+\tau_{\rm self}}.
\end{equation}
The gravitational condensation time is given by 
\begin{equation}
    \label{eq:taugrav}
\tau_{\gr}   = \frac{b}{48\pi^3}\frac{m_a^3 v^6}{G^2 \rho_a^2\log \left(m_a v R\right)}~,
\end{equation}
and the self-interaction condensation time is 
\bea
\label{eq:tauself}
\tau_{\rm self}=\frac{64 d m_a^7 v^2}{3\pi \rho_a^2 \lambda^2}~.
\eea

The parameters $b,d\sim\mathcal{O}(1)$ are numerical coefficients extracted from numerical simulations \cite{Chen:2020cef}. In this work, we focus on the scenario where the axion self-coupling strength is given by $\lambda=m_a^2/f_a^2$. These are standard formulas that calculate the relaxation scale of axion gas with a given velocity distribution and agree surprisingly well with the axion star formation timescale found in numerical simulations \cite{Levkov:2018kau,Eggemeier:2019jsu,Chen:2021oot,Chan:2022bkz}.
The velocities and densities should be calculated from halo parameters. Since the halo central density is higher and the velocity is smaller than the halo outskirts, the axion star formation rate will be enhanced compared to the halo outskirts. In the post-inflationary scenario of axions, axion miniclusters will form in early times and evolve with time \cite{Xiao:2021nkb}, greatly enhancing the axion star formation rate subject to the disruption effect of axion miniclusters in galaxies \cite{Kavanagh:2020gcy,Shen:2022ltx}. Here we do not consider the enhancement from axion miniclusters but focus on axion star formation in the standard massive dark matter halos.
We considered the halo density in the region where the enclosed dark matter mass is the same as the black hole mass, which characterizes the halo environment in the vicinity of supermassive black holes. The detailed calculations are presented in \Appx{accrete}. The characteristic mass scale of the axion star can be estimated by equating the virial velocity of the axion star and the halo velocity, which gives \cite{Levkov:2018kau,Eggemeier:2019jsu}
\begin{equation}
\label{eq:M_star_overline}
    \overline{M}_{*} \simeq \frac{3 \,v}{G m_a}
\end{equation}
where $v$ is the velocity of axion waves in dark matter halos. 
The mass growth of axion stars is found to exhibit a power-law growth after the initial thermalization
\begin{equation}
\label{eq:M_star_evo}
    M_{*}(t)=  \overline{M}_{*} \left(\frac{t}{\tau}\right)^{1/2},
\end{equation}
where $\tau$ is the condensation timescale given in \Eq{tau}. 
It is still debatable if such power-law growth can be extrapolated to masses $M\gg \overline{M}_{*}$ \cite{Eggemeier:2019jsu,Dmitriev:2023ipv}.  However, for the parameter space of interest in this paper, the axion stars we are considering are very light and within the mass range where the power-law growth is applicable. Since the ground state of the gravitational atom can be absorbed by black holes, one should also include the decay time, $t_{\rm decay}$, of axion star solutions near black holes \cite{Detweiler:1980uk,Baryakhtar:2017ngi,Baumann:2019eav}. The lifetime of axion stars in the ground state is much shorter than that in excited states. Since we do not have a population analysis for the occupancy number of different eigenstates, we assume all the axion stars are in the ground state, which will be the conservative limit. 
If the lifetime of ground state $t_{\rm decay}$ is much shorter than Hubble time $t_H$, one could determine the axion star mass using
\bea
\label{eq:M_star_today_main_text}
M_* \simeq \overline{M}_{*}\,\left(\frac{t_{\rm decay}}{\tau}\right)^{1/2}\,.
\eea
If $t_H$ is smaller than $t_\decay$, we replace $t_\decay$ in \Eq{M_star_today_main_text} by $t_H$. It is worth emphasizing again that we use the standard formation rate for the axion stars in the Galactic Center, which should be a conservative calculation. The axion star itself can help capture more axions \cite{Budker:2023sex} and the BH may enhance the accretion rate of axions.

\section{Axion Sensitivities from black hole polarimetry}\label{sec:sensitivity}

Now we will apply our previous calculations to obtain the sensitivity of black hole polarimetry to axion dark matter. EHT observations have imaged not only the shadows of supermassive black holes such as M87 and Sgr A$^*$ on event-horizon scales~\cite{EventHorizonTelescope:2019ggy,EventHorizonTelescope:2019pgp,EventHorizonTelescope:2022wkp}, but also their polarized emissions \cite{EventHorizonTelescope:2021srq,EHT:2024nwx}. The polarization of emission can be quantized by the complex linear polarization field, and the phase defines the EVPA. A large axion field value will cause an oscillating angle, which can be constrained by the time variability of the measured EVPA. For SMBHs, the photon rings and the jets are good polarization sources, and measurable EVPA with time modulation will be generated as the emitted lights go through the axion stars around SMBHs. Because the period of the time modulation of the axion field is longer than the axion star size given, $G M_\text{BH} m_a < 1$, the washout effect is negligible.

Using the axion star mass, one can derive the sensitivity on $\ga$ with the formula in \Eq{theta_polar_approx}. The axion parameter space that can be probed by the polarimetric measurements of M87$^*$ and Sgr A$^*$ is presented in \Fig{ma_ga_plt}. The sensitivities of EHT observations on the polarization angle rotation are taken as 3$^\circ$\cite{Johnson:2015iwg}, which have been achieved in observations of Sgr A$^*$ with an exposure of tens of minutes. The ngEHT should provide even better sensitivity to the oscillating polarization angle, which is as small as $0.05^\circ$~\cite{Ayzenberg:2023hfw,Wang:2023eip}. M87$^*$ generally probes light axion masses that correspond to a longer oscillating period, allowing for a longer sampling time and better sensitivity. Therefore, 3$^\circ$ precision can be considered as a conservative estimate, and we leave more dedicated data analysis for future explorations. It is also important to note that the axion densities around the SMBHs from the axion star formations are much larger than the axion density during the CMB epoch, from which we know that the axion field value around SMBHs is much larger. Therefore, the black hole polarization measurements through EHT and ngEHT are much more sensitive than the CMB measurements. For detailed discussions, one can refer to \Appx{axion_density_smbh}.

There are two competing effects that determine the sensitivities. In general, a large star formation rate will lead to heavier critical axion stars, which provides better sensitivities to lighter axions due to the large Bose enhancements. The decay of axion stars near black holes should also be taken into account, which suppresses the sensitivities when $GM_{\rm BH}m_a$ is close to 1. On the other hand, a larger value of $GM_{\rm BH}m_a$ corresponds to more enhancements on the axion field value of critical stars, which tends to make the sensitivities better. Therefore, we see the sensitivity is almost flattened in a wide range of axion masses, as shown in \Fig{ma_ga_plt}. Note that the axion decay constant $f_a$ is also a crucial part of the story to calculate the axion parameter space that can be probed. 
A large self-interaction can enhance the formation rate of axion stars, which can expand the discovery space for axion parameters until the critical mass of axion stars is reached, as shown in \Fig{ma_ga_plt}. But axion star formation can happen even without self-interactions, which will give us the most conservative estimation in the sensitivity to axion parameters from black hole polarimetry, as shown in the shaded region of \Fig{ma_ga_plt}.

Black hole polarimetry has to resolve the size of the axion star to probe the polarization rotation induced by axion stars.
It is challenging to resolve the near-horizon region of supermassive black holes if they are at high redshifts. The required angular resolution to probe axions is given by $\sim R_{*}/D_s$, where $R_*$ is the axion star radius given in \Eq{R_star} and $D_s$ is the comoving distance of the supermassive black hole. Space interferometry in the future will provide a better angular resolution with longer baselines \cite{Gurvits:2022wgm}, which can resolve axion stars near supermassive black holes much more precisely. Therefore, we expect polarimetric imaging of the near-horizon region of supermassive black holes will also serve as competitive axion searches in the future, with great potential to discover axion dark matter.

\section{Conclusion}\label{sec:conclusion}
 In this work, we systematically study the axion star configuration, axion star formation rate, and axion birefringence effect around supermassive black holes and propose black hole polarimetry as sensitive axion dark matter detectors. The axion birefringence effect is distinguishable from astrophysical effects on the rotation of polarization angle since it is frequency-independent and modulating at the timescale of $1/m_a$ coherently over the size of an axion star. Therefore, if a large axion field value is developed around supermassive black holes, it will become a potential discovery machine for axion dark matter. 
 In our study, a new stable solution of axion stars in the gravitational background of black holes is obtained, and the axion star mass is calculated based on the condensation timescale of axion waves considering both gravitational and axionic self-couplings, showing that a large axion field around black holes is indeed possible for a broad range of parameter space. The maximal mass of axion stars is determined by the axion self-coupling, where a stronger self-coupling corresponds to a smaller maximal mass and larger axion star formation rate. We studied the axion star formation rate using analytical models motivated by numerical simulations. The axion star formation rate from gravity is already significant, and the self-coupling of axions can further enhance it, expanding the parameter space that can be probed. When the maximal mass of axion stars is achieved with a sufficiently large axion self-coupling but not too strong to reduce the maximum mass, more axion parameters can be probed.
 The exact law for the growth of the mass of axion stars is still being actively studied and future numerical studies might change the axion star mass predicted in this work. However, our calculation is conservative as we only considered the condensation of axion waves in dark matter halos, which has been extensively studied in the literature. We did not include the possible enhancements in the formation rate from the gravitational potential of black holes. 
 We show that polarimetric imaging of the near-horizon region of supermassive black holes which can resolve the axion star profile will provide sensitive probes to axion-photon couplings. . 
 Future observations with the next-generation Event Horizon Telescope will greatly expand the axion mass range it can probe with more sources and improve the sensitivity on the axion-photon coupling constant $\ga$ with better measurements on the polarization angle. Black hole polarimetry will not only be interesting in astrophysics but also provide one of the best complimentary searches on the nature of dark matter.

\section*{Acknowledgements}

We thank Patrick J. Fox, Yifan Chen, Joshua Eby, and Michael A. Fedderke for their helpful discussions and comments on the draft. XG is supported by James Arthur Graduate Associate (JAGA) Fellowship. HX is supported by Fermi Research Alliance, LLC under Contract DE-AC02-07CH11359 with the U.S. Department of Energy. The work of LTW is supported by DOE grant DE-SC-0013642. This work was performed in part at the Aspen Center for Physics, which is supported by National Science Foundation grant PHY-2210452.

\appendix

\section{Axion Star Configuration in SMBH Background}\label{appx:astar_config}

In this section, we solve the axion star configuration in the SMBH background. In the SMBH background, the axion field obeys the Klein-Gordon in curved spacetime, which is written as
\bea
\label{eq:a_KG}
\frac{1}{\sqrt{-g}} \partial_\mu(\sqrt{-g} g^{\mu \nu} \partial_\nu a) = - V'(a).
\eea
The metric of the SMBH background in the Newtonian limit is
\bea
\label{eq:metric}
g_{\mu \nu} = \text{diag}\big( 1+2\Phi, - (1-2\Phi), -r^2, -r^2 \sin^2\theta \big), 
\eea
where $\Phi$ represents the gravitational potential. In \Eq{metric}, we use the Newtonian approximation because we focus on the region where $r\gg GM_\BH$. Neglecting the axion self-interaction, doing the time average, and neglecting the higher order terms, we write \Eq{a_KG} as
\bea
\label{eq:a_schrodinger}
\pmb{\nabla}^2 \Theta = 2 \left(m^2_a \Phi+\frac{m^2_a -\omega^2_a}{2} \right) \Theta,
\eea
where the spatial Laplace operator can be written as
\bea
\pmb{\nabla}^2 = \frac{\partial^2}{\partial r^2} + \frac{2}{r} \frac{\partial }{\partial r}
\eea
because the axion's ground state configuration is spherically symmetric. 

In the Newtonian limit, the gravitational potential obeys the Poisson equation, i.e., the linearized Einstein equation. Here, we have 
\bea
\label{eq:Poisson}
\pmb{\nabla}^2 \Phi = 4 \pi G \rho_\tot,
\eea
where
\bea
\label{eq:rho}
\rho_\tot = \rho_* + \rho_{\rm BH}.
\eea
Here, $\rho_*$ denotes the density of the axion star and $\rho_\BH$ denotes the density of the black hole. Because we focus on the region where $r \gg G M_\BH$, we  approximately have
\bea
\rho_{\rm BH} \simeq M_\BH \delta^{(3)}(\pmb{r}).
\eea
Because we focus on the situation where $M_\BH \gg M_*$, after solving \Eq{Poisson}, we have 
\bea
\label{eq:Phi_BH}
\Phi \simeq - \frac{G M_\BH}{r}. 
\eea
After substituting \Eq{Phi_BH} and imposing the boundary condition $\Theta(r\rightarrow \infty)=0$, we get the ground state solution
\bea
\label{eq:Theta_sol}
\Theta(r) = \Thetazero \exp\left( - G M_\BH m_a^2 r \right)
\eea
with the eigenfrequency
\bea
\label{eq:omega_sol}
\omega_a = m_a \sqrt{ 1 - \left(G M_\BH m_a \right)^2 }. 
\eea
From \Eq{Theta_sol}, we know that the radius of the axion star is $R_* \sim 1/G M_\BH m_a^2$. When $r \gg R_*$, the axion field decreases exponentially. To be more specific, we define $R_*$ to be radius which envelopes $99\%$ of the axion star mass, then we have 
\bea
\label{eq:R_star_99}
R_* = \frac{c_{99}}{G M_\BH m_a^2}, \quad \text{where $c_{99} \simeq 4.2$}. 
\eea
From \Eq{omega_sol}, we find that to have a stable axion configuration, we need 
\bea
m_a \leq \frac{1}{G M_\BH}.
\eea
Otherwise, the axion's oscillation frequency is imaginary, which means that the axion is absorbed by the black hole when its size is comparable with the black hole's event horizon.

\section{Axion Star Energy in SMBH Background}\label{appx:axion_star_energy}

In this section, we will do a detailed calculation of the axion star's energy in the SMBH background. The axion's energy density can be written as 
\bea
\label{eq:rho_star_average}
\rho_* = \frac{\omega^2_a \fa^2\Theta^2}{4}  + \frac{\fa^2}{4}  \left(\frac{d \Theta}{dr}\right)^2 + \frac{m_a^2 \fa^2 \Theta^2}{4} - \frac{\lambda \fa^4 \Theta^4}{64},
\eea
where we do the time-average over the time-oscillation terms. In \Eq{rho_star_average}, the first term denotes the oscillation energy density, the second term denotes the gradient energy density, the third term denotes the potential energy density from axion's mass term, and the fourth term denotes the potential energy density from axion's self-interaction. 

Based on the discussion above, the total energy of axion star can be written  as
\bea
\label{eq:M_star_tot}
E^\tot_* = E^\osc_* + E^\grad_* + E^\mass_*  + E^\lambda_* + E^\grav_*.
\eea
Each term in \Eq{M_star_tot} can be written as
\begin{equation}
\label{eq:M_star_tot_calc}
\left\{
\begin{aligned}
&E_*^\osc 
=\int d^3 \pmb{r}  \, \frac{\omega_a^2f_a^2 \Theta^2}{4} = \frac{\pi \fa^2 \omega_a^2 \Theta_0^2}{ 4 G^3 m_a^6 M_\BH^3 }\\
&E_*^\grad = \int d^3 \pmb{r} \, \frac{\fa^2}{4} \left(\frac{d \Theta}{dr} \right)^2  = \frac{\pi \fa^2 \Thetazero^2}{ 4 G m_a^2 M_\BH }\\
&E_*^\mass = \int d^3 \pmb{r} \, \frac{m_a^2 \fa^2 \Theta^2}{4} = \frac{\pi \fa^2 \Thetazero^2}{4 G^3 m_a^4 M_\BH^3}\\
&E_*^\lambda = - \int d^3 \pmb{r} \, \frac{\lambda \fa^4 \Theta^4}{64} = - \frac{\pi \lambda \fa^4 \Thetazero^4 }{512 G^3 m_a^6 M_\BH^3}\\
&E_*^\grav = - \int d^3 \pmb{r} \, \frac{G M_\BH \rho_*}{r} \simeq - \frac{\pi \fa^2 \Thetazero^2}{ 2 G m_a^2 M_\BH }
\end{aligned}\right. .
\end{equation}

When $GM_\BH m_a \ll 1$, $\omega_a \simeq m_a$; therefore, we have 
\bea
\label{eq:M_star_result}
M_* \simeq E^\osc_* + E^\mass_* \simeq \frac{\pi \Thetazero^2 \fa^2}{ 2 G^3 M_\BH^3 m_a^4 }, 
\eea
where contributions from $E^\grad_*$ and $E^\lambda_*$ are negligible. Representing $\Theta_0$ in terms of $M_*$ using \Eq{M_star_result}, we can write the gradient energy listed in \Eq{M_star_tot_calc} as
\bea
\label{eq:E_star_grad_calc}
E^\grad_* \simeq \frac{G m_a^2 M_\BH^2 M_*}{2}. 
\eea
Similarly, we can write the self-interaction energy listed in \Eq{M_star_tot_calc} as 
\bea
\label{eq:E_star_lambda_calc}
E^\lambda_* \simeq \frac{\lambda\, G^3 m_a^2 M_\BH^3 M_*^2}{128 \pi},
\eea
and
the gravitational energy listed in \Eq{M_star_tot_calc} as
\bea
\label{eq:E_star_grav_calc}
E_*^\grav \simeq -G^2 m_a^2 M_\BH^2 M_*. 
\eea

Substituting \Eq{M_star_result} into the equation $\Delta\phi_\polar \simeq \ga \Thetazero \fa/2$, we have
\bea
\label{eq:theta_polar}
\Delta\phi_\polar \simeq \ga \left(\frac{G^3 M_\BH^3 m_a^4 M_*}{2\pi}\right)^{1/2}.
\eea
Substituting \Eq{M_star_max_2} into \Eq{theta_polar}, we can write the rotation angle of the light polarization at the critical mass as
\bea
\label{eq:phi_polar_max_2}
\Delta\phi^{\max}_\polar \simeq \frac{4}{\sqrt{3}} \,\ga \fa \, G M_\BH m_a. 
\eea
Therefore if axions with larger $f_a$ can form axion stars at critical masses, the polarization rotation angle will be larger.

\section{Critical Mass of the Axion Star in SMBH Background}\label{appx:critical mass}

In this section, we derive the axion star's critical mass $M_*^{\max}$ in the SMBH background with $M_*\ll M_\BH$ based on the variation principle.  Here we have the energy of the axion star, i.e., 
\bea
\label{eq:E_star}
E_* =  c_1 \frac{M_*}{2m_a^2 R_*^2} - c_2\frac{\lambda M_*^2}{12 m_a^4 R_*^3} - c_3 \frac{G M_* M_\BH}{R_*},
\eea
where the first term is the gradient energy, the second term is the energy from the axion self-interaction, and the third term is the gravitational energy of the axion star in the SMBH background. Comparing \Eq{E_star} with \Eq{R_star_99}, \Eq{E_star_grad_calc}, \Eq{E_star_lambda_calc}, and \Eq{E_star_grav_calc}, we can determine that 
\bea
\label{eq:c_123_SMBH}
c_1 = c_{99}^2, \, c_2 = \frac{3 c_{99}^3}{32 \pi}, \, c_3 = c_{99}.
\eea
It is worth noting that the numerical values of $c_1$, $c_2$ in \Eq{c_123_SMBH} are different from those of self-gravitating axion stars where $c_1 \simeq 9.9$, $c_2 \simeq 0.85$ \cite{Visinelli:2017ooc, Ruffini:1969qy,Membrado:1989ke}. 

Doing the variation of \Eq{E_star} and imposing the condition $dE_*/dR_* =0$ for the configuration's stability, we have
\bea
\label{eq:R_star_vary}
R_* = \frac{c_1}{2 c_3 G M_\BH m_a^2}\left[1 \pm \sqrt{1-\frac{c_2 c_3}{c_1^2} \lambda G M_* M_\BH  } \right].
\eea
In the weak coupling limit where $\lambda$ is negligible, we have $R_* \simeq c_{99}/GM_\BH m_a^2$. This is consistent with the result in \Eq{R_star_99}. However, the axion star is unstable and starts to explode if $M_*$ becomes too large. If the axion star stays in a stable configuration, $R_*$ in \Eq{R_star_vary} has to be a real number. This requires that 
\bea
M_* \leq M_*^{\max} = \frac{32 \pi}{3 \lambda GM_\BH}.  
\eea
During the axion accretion around the SMBH, the axion star's mass grows as $M_* \propto t^{1/2}$ until $M_* \simeq M_*^{\max}$. Afterward, the axion star triggers the bosenova and maintains a mass $\sim M_*^{\max}$. After the bosenova, the axion field value is only changed by an $\mathcal{O}(1)$ factor, so the conclusion will not be changed qualitatively.

Before ending this section, we want to comment on the different $M_*\text{-}R_*$ relation for the axion star in the SMBH background and the self-gravitating axion star. As shown in \Eq{R_star_vary}, the black hole mass replaces the axion star mass in the usual axion star mass-radius relation. This is easily understandable because the gravitational potential of axion stars is more important in self-gravitating axion stars while the gravity of black holes is dominating in our case. Therefore, the gravitational potential of black holes naturally holds the axion stars, and axion stars become more compact from the gravitational pull of black holes.

\section{Axion Accretion in SMBH Background}\label{appx:accrete}

\begin{figure*}[t]
\centering
\includegraphics[width=1.2\columnwidth]{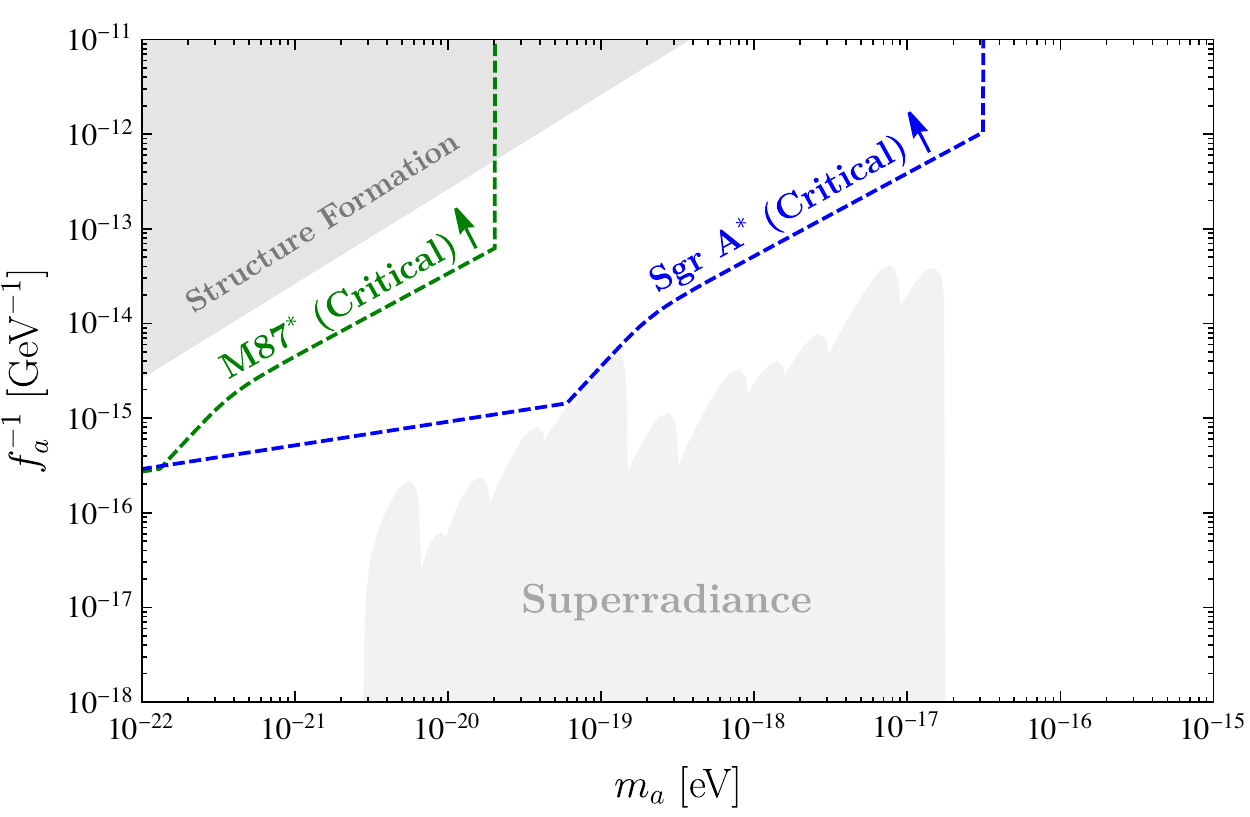}
\caption{
The axion parameter space that leads to the formation of axion stars at critical masses near the supermassive black holes is plotted. A smaller $f_a$ is ideal for forming critical stars because the formation rate of axion stars is enhanced with a small $f_a$ and the corresponding critical mass of axion star is also small. However, it is not required to obtain the sensitivity since axion stars can also form with gravitational interactions.
We again considered axion star formation near both M87$^*$ and Sgr A$^*$. The light gray region is the axion parameter space excluded by the superradiance of SMBHs \cite{Baryakhtar:2020gao,Mehta:2020kwu,Unal:2020jiy}. Ultralight axion dark matter with masses lighter than $10^{-21}\,\eV$ is excluded by  Lyman-$\alpha$ forest \cite{Irsic:2017yje, Kobayashi:2017jcf, Armengaud:2017nkf, Zhang:2017chj, Nori:2018pka, Rogers:2020ltq}. The ultrafaint dwarf galaxies may also place a limit on the dark matter mass to be heavier than $10^{-19}\eV$~\cite{Dalal:2022rmp}. The gray region on the upper left corner shows the constraints from structure formation \cite{Arvanitaki:2014faa, Fan:2016rda, Cembranos:2018ulm, Budker:2023sex}.  } 
\label{fig:ma_fa}
\end{figure*}

In this section, we discuss the accretion of the axion within the background of the SMBH. Axion mass accumulates around the SMBH, obeying power-law growth. This effect leads to the mass growth of the axion star. In the meantime, because the part of the axion star inside the horizon of the SMBH is absorbed, it experiences exponential decay. Considering these two competing effects, we have
\bea
\label{eq:Mstar_Eq}
\frac{d M_*}{dt} = \frac{\overline{M}^2_{*}}{\tau} \frac{1}{M_*} - M_* \Gamma_{\text{decay}},
\eea
where $M_*$ begins to evolve from zero to nonzero values at $t=0$. In the right-hand side of \Eq{Mstar_Eq}, the first term represents the accumulation of axions obeying the power law $M_*(t) \propto t^{1/2}$. $\tau$ represents axion's relaxation time including both the gravitational interaction and the self-interaction. $\overline{M}_*$ is the characteristic mass of the axion star. The second term represents the exponential decay of the axion star in the gravitational field of SMBH, which appears as the imaginary part of the energy eigenvalue. From \cite{Detweiler:1980uk,Baryakhtar:2017ngi,Baumann:2019eav}, we know that the decay rate is 
\bea
\label{eq:Gamma_Decay}
\Gamma_{\text{decay}} \sim (m_a \Omega_\BH- \omega_a) \times (G M_\BH m_a)^{5+4\ell},
\eea
where $\Omega_\BH$ is the spin of the black hole, $\ell$ is the angular momentum quantum number, and $\omega_a$ is the real part of the axion oscillation frequency $\omega_a = m_a \sqrt{1 - \left( G M_\BH m_a \right)^2 }$. Because we focus on the region where $m_a \ll 1/G M_\BH$, there is $\omega_a \simeq m_a$. For the axion star in the ground state, we have $\ell=0$; therefore, the inverse of the decay time is
\bea
\label{eq:Gamma_Decay_l0}
t_{\text{decay}}^{-1} \sim  m_a \times  (G M_\BH m_a)^5. 
\eea
During the evolution of the Universe, the mass of an axion star grows over time with a power law until the two terms in the left-hand side of \Eq{Mstar_Eq} cancel each other. Given that the age of the Universe is $t_H \simeq 1/H_0$, today's axion star mass is
\bea
\label{eq:M_star_today}
M_* = \overline{M}_* \times \left( \frac{\min(t_\decay,t_H)}{\tau} \right)^{1/2}
\eea
before $M_*$ reaches the critical mass $M_*^{\max}$. 

To calculate $\tau$ in \Eq{M_star_today}, we need to know $\rho_a$. To specify this, we choose $\rho_a$ to be the energy density at $r=r_0$, where $r_0$ satisfies
\bea
\label{eq:M_halo_BH_eq}
M_\halo(r<r_0) \simeq M_\BH. 
\eea
In the region where $r > r_0$, the SMBH gravitational field is subdominant, and therefore the axions in this region do not feel the existence of the SMBH. Moreover, because the axion accretion time mostly depends on the outside mass shell, using $\rho_a$ at $r=r_0$ is a conservative approach.

We assume that the axion halo has the NFW profile
\bea
\rho_\NFW(r) = \frac{\rho_s}{\frac{r}{r_s} \left( 1 + \frac{r}{r_s} \right)^2 }.
\eea
If the halo mass integration is cut off at $r_{200}$, where $r_{200}$ is the radius where the averaged halo density is $200$ times the average matter density $\rho_m$, we have
\bea
\label{eq:rho_s_r_s}
\rho_s = \frac{ \Delta_{200} \rho_m c^3}{3 f(c)} \,\,\,\,\, \text{and}\,\,\,\,\,  r_s = \left(\frac{M_\halo}{ 4 \pi \rho_s f(c) }\right)^{1/3},
\eea
where
\bea
\label{eq:fc}
f(c) = \log(1+c) - \frac{c}{1+c},
\eea
and $\Delta_{200} = 200$. Here, $c$ is the dark matter halo concentration factor and is generally chosen to be $c\sim 10$. To solve $r_0$, we substitute \Eq{rho_s_r_s} and \Eq{fc} into \Eq{M_halo_BH_eq} and then do the Talyor expansion over $r_0/r_s$ given that $r_0 \ll r_s$. After some algebra, we have
\bea
\label{eq:r0}
r_0 \simeq \left( \frac{M_\BH}{2 \pi r_s \rho_s} \right)^{1/2}.
\eea
We also have
\bea
\label{eq:v0}
v_0 = \left(\frac{2 G M_\BH}{r_0}\right)^{1/2}
\eea
and
\bea
\label{eq:rho0}
\rho_0 = \frac{\rho_s}{\frac{r_0}{r_s} \left( 1 + \frac{r_0}{r_s} \right)^2 }. 
\eea
Substituting \Eq{rho0} and \Eq{v0} into the formula of the condensation timescale, we will be able to calculate the axion star formation rate. For example, the condensation timescale caused by self-interactions can be expressed as
\bea
\label{eq:tau_0}
\tau_{\rm self} \simeq \frac{64 d m_a^3 f_a^4 v^2_0}{3\pi \rho_0^2}.
\eea
The gravitational condensation time is given by 
\begin{equation}
    \label{eq:taugrav_0}
\tau_{\gr}  \simeq \frac{b}{48\pi^3}\frac{m_a^3 v^6_0}{G^2 \rho^2_0\log \left(m_a v_0 R\right)}~.
\end{equation}
Based on \Eq{M_star_today} and \Eq{taugrav_0}, we can explain the power-law behavior of the solid lines in \Fig{ma_ga_plt}, where the gravitational condensation dominates. When $m_a$ is small, the decay rate is suppressed, which leads to $t_H < t_\decay$. Thus we have $M_* \simeq \overline{M}_* (t_H/\tau_\gr)^{1/2}  \propto m_a^{-5/2}$. Because $\rho_* \sim M_*/R_*^3$, we have $\Delta \phi_\polar \propto \ga m_a^{3/4}$. Given the determined $\Delta \phi_\polar$, we conclude that the solid contours obey $\ga \propto m_a^{-3/4}$. When $t_\decay < t_H$, we have $M_* \simeq \overline{M}_* (t_\decay/\tau_\gr)^{1/2}$. From \Eq{Gamma_Decay_l0}, we have $t_\decay \propto m_a^{-6}$, which leads to $M_* \propto m_a^{-11/2}$. Therefore, we have $\Delta \phi_\polar \propto \ga m_a^{-3/4}$, from which we have $\ga \propto m_a^{3/4}$. 

It is worth noting that the gravitational interaction will dominate over self-interaction if $f_a>M_{\rm pl}\, v$ during the formation of axion stars. Also, note that the power-law growth $M_{*}(t)\propto t^{1/2}$ is only a benchmark model, and the power-law index may be different. For example, the halo profile may contribute to the power-law growth. If a star is formed within a small radius of the halo, which is the situation here, the mass contained in this region is small. For an NFW profile, the mass contained within $r$ is
\bea
M(r)|_{r\rightarrow 0}=4\pi \rho_s r_s^3 f(r/r_s) \simeq 2\pi \rho_s r_s r^2.
\eea
At a small radius, the mass enclosed will form an axion star when all the axion waves are scattered with each other and thermalized.
The formation timescale dominated by self-interactions is $\tau_{\rm self}\propto v^2/\rho^2$. At a small radius of an NFW halo, the dark matter density and velocity scale as $\rho\propto 1/r$ and $v\propto \sqrt{r}$, respectively. Therefore, $M(t)\propto r^2\propto t^{2/3}$. Similarly, if gravity dominates the axion star formation, $\tau_{gr}\propto v^6/\rho^2\propto r^5$ at a small radius and the mass growth power law is given by $M(t)\propto t^{2/5}$. One may use a different power-law index for the growth behavior of axion stars but this should not change the result significantly.

In \cite{Budker:2023sex}, the growth of axion stars is further enhanced by the axion star itself since it will capture axions that are originally unbound, triggering exponential growth on the axion star mass before it reaches the critical mass. Such a capture rate may not be relevant in our case since the dark matter environment is so dense and the scattering of axion waves is already significant enough to form axion stars. As shown in \Eq{Mstar_Eq}, the growth term from external axion scattering is $\overline{M}_{*}^2/(\tau {M_*})$ while axion star capture will introduce a new growth term $M_*/\tau$, which is subdominant when $M_*<\overline{M}_*$.

In \Fig{ma_fa}, we plot the $m_a\text{-}f_a$ plane that leads to the formation of critical stars near supermassive black holes. Both $m_a$ and $f_a$ are relevant for the formation and growth of axion stars as well as their critical mass. When the critical mass is reached, the sensitivity to axion-photon couplings is maximized. It is worth noting that our calculation has been conservative because we considered the decay of axion stars near black holes using the decay rate of axion stars in the ground state. The large decay rate of axion stars when $GM_{\rm BH}m_a\sim 1$ is the reason why a small $f_a$ is needed to form critical stars at the high mass end.
Future simulations of axion fields might be able to determine the exact axion state distributions and we will have a better understanding of the axion parameter space that leads to the formation of critical stars.

\section{Axion Densities Around SMBHs}\label{appx:axion_density_smbh}

In this section, we numerically estimate the axion densities developed through the axion accretion around the SMBHs and compare them with the axion densities during the CMB epoch~($z_\CMB \sim 10^3$). Through this, we can intuitively reveal the reason why the axion constraints from the black hole polarimetry are stronger than the CMB polarization measurements. Here, we take Sgr A$^*$ as an example and the discussion of M87$^*$ is similar. From \Eq{R_star}, the radius of the axion star around Sgr A$^*$ is
\bea
\label{eq:R_star_SgrA_num}
R_*|_\text{Sgr A$^*$} \sim 10^{-4}\,\text{pc} \left(\frac{10^{-18}\eV}{m_a}\right)^{2}. 
\eea
To simplify the discussion, we assume that the axion star formation merely comes from the gravitational interaction. Utilizing \Eq{M_star_overline}, \Eq{M_star_today_main_text} and \Eq{R_star_SgrA_num}, we have the density of the axion star around Sgr A$^*$, which is
\bea
\label{eq:ruo_star_SgrA_num}
\rho_*|_\text{Sgr A$^*$} \sim 10^{11} \,\gev/ \cm^{3} \left( \frac{m_a}{10^{-18}\eV}\right)^{1/2}. 
\eea
To compare the sensitivities from the EHT with the one through the CMB polarization, we also calculate the axion density at the last scattering surface using $\rho_a|_\text{CMB}\sim  (1+z_\CMB)^3 \overline{\rho}_{a,0}$. Here, $\overline{\rho}_{a,0} \sim 1 \, \kev/\cm^3$ is the average density of the axion dark matter today. Therefore, we have
\bea
\label{eq:ruo_CMB_num}
\rho_a|_\CMB \sim 10^3 \, \gev/\cm^3.
\eea
\\
Noticing that \Eq{ruo_star_SgrA_num} is much larger than \Eq{ruo_CMB_num}, we know that the axion field values around the SMBHs are much larger than the axion field value at the last scattering surface. Because the sensitivity of Planck 2018 is around $\Delta \phi_{\rm polar} \sim 1^{\circ}$ and the future CMB measurement can only improve by less than two orders of magnitude~\cite{Fedderke:2019ajk,Diego-Palazuelos:2022dsq}, we prove that the black hole polarization measurement is much more sensitive to detect the axion dark matter compared with CMB polarization measurement.

\bibliographystyle{apsrev4-2}
\bibliography{ref}
\end{document}